\documentclass[conference]{IEEEtran}
\pdfoutput=1

\usepackage[T1]{fontenc} 
\usepackage[utf8]{inputenc}
\usepackage{tikz}
\usepackage{pgfplots}
\pgfplotsset{compat=1.8}
\usepackage{pgfplotstable}
\usepgfplotslibrary{groupplots}
\usepackage{afterpage}
\usepackage{siunitx}
\usepackage{url}
\usepackage{amsfonts}
\usepackage{amsmath}
\usepackage[caption=false,font=footnotesize]{subfig}
\usepackage{todonotes}
\usepackage{graphicx}
\usepackage{multirow}
\usepackage{booktabs}
\usepackage{hyperref}
\usepackage{natbib}
\usepackage{balance}
\usepackage{url}

\setcitestyle{numbers,square,comma}

\usetikzlibrary{arrows,automata}
\usetikzlibrary{positioning}

\tikzset{
	state/.style={
		rectangle,
		rounded corners,
		draw=black, very thick,
		minimum height=2em,
		inner sep=2pt,
		text centered,
	},
}

\graphicspath{{./figures/}}

\begin{document}
%
\title{Describing Subjective Experiment Consistency\\by $p$-Value P--P Plot}

\author{\IEEEauthorblockN{Jakub Nawała\IEEEauthorrefmark{1},
Lucjan Janowski\IEEEauthorrefmark{1}, Bogdan Ćmiel\IEEEauthorrefmark{2} and
Krzysztof Rusek\IEEEauthorrefmark{1}}
\IEEEauthorblockA{\IEEEauthorrefmark{1}AGH University of Science and
Technology\\Department of Telecommunications\\Kraków, Poland\\jnawala@agh.edu.pl}
\IEEEauthorblockA{\IEEEauthorrefmark{2}AGH University of Science and Technology\\
Department of Mathematical Analysis,\\
Computational Mathematics and Probability Methods\\
Kraków, Poland}
}

\maketitle

\begin{figure*}[t!]
    \centering
    \subfloat[]{
        \includegraphics[height=4.3cm]{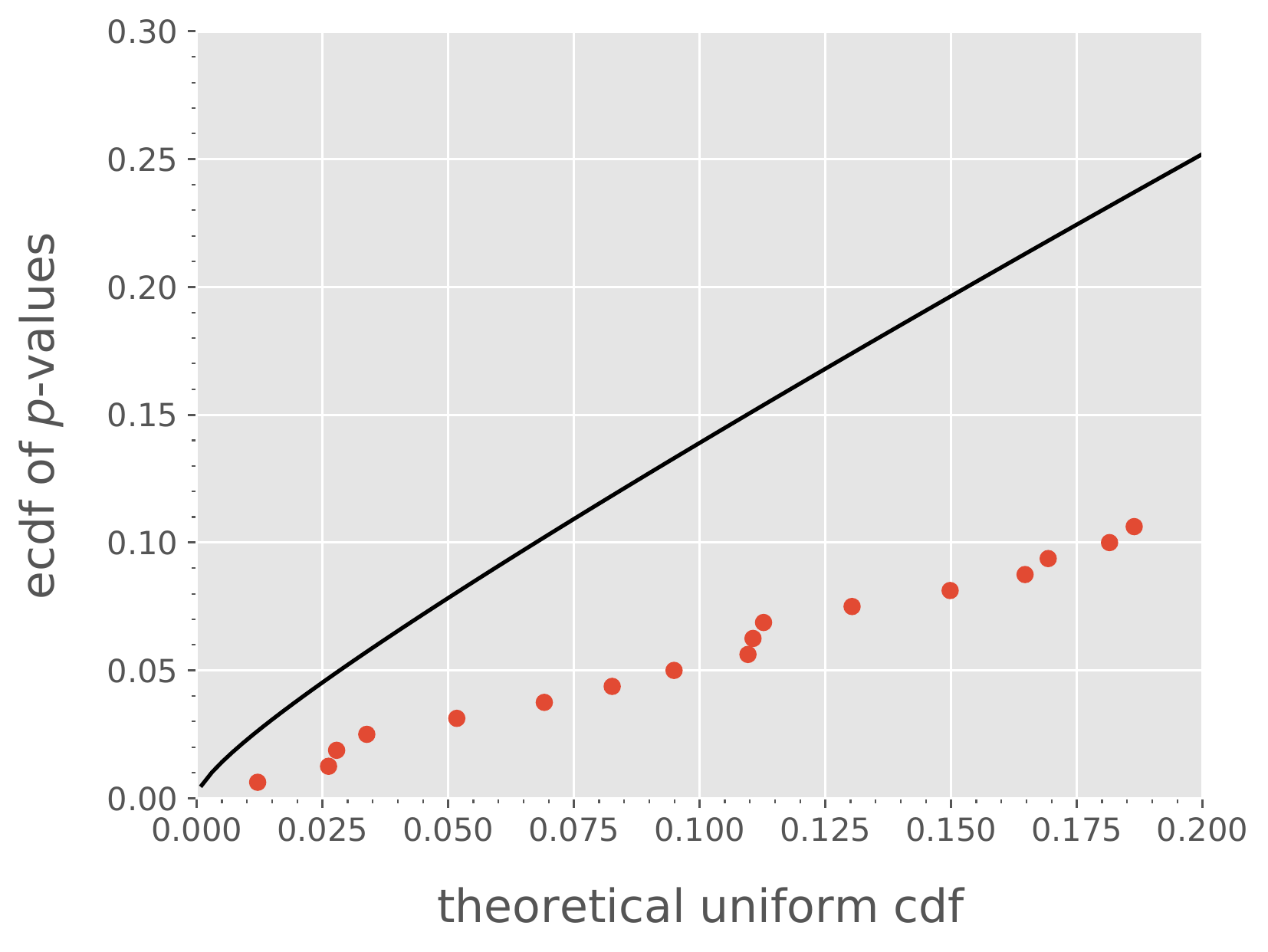}
        \label{fig:ex_consistent_exp}
        \hfill
    }
    \subfloat[]{
        \includegraphics[height=4.3cm]{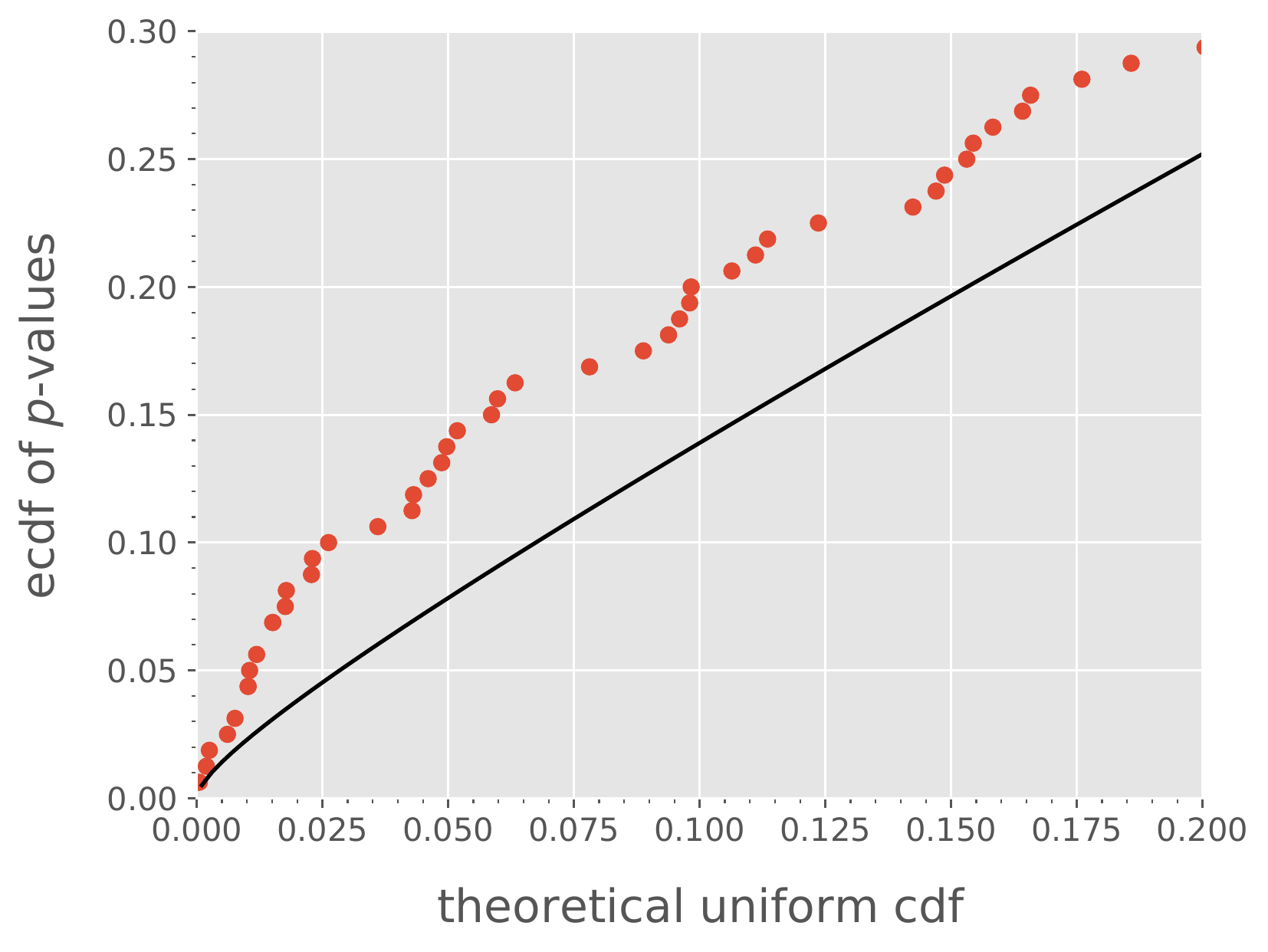}
        \label{fig:ex_inconsistent_exp}
        \hfill
    }
    \caption{Our main contribution is a tool allowing to discern
	between consistent (a) and inconsistent
	(b) subjective experiment.}
	\label{fig:teaser}
\end{figure*}

\begin{abstract}
\boldmath
There are phenomena that cannot be measured without subjective testing.
However, subjective testing is a complex issue with many influencing
factors. These interplay to yield either precise or incorrect results.
Researchers require a tool to classify results of subjective experiment as 
either
consistent or inconsistent. This is necessary in order to decide whether to 
treat the
gathered scores as quality ground truth data.
Knowing if subjective scores can be trusted is key to drawing valid
conclusions and building functional tools based on those scores (e.g.,
algorithms assessing the perceived quality of multimedia materials).
We provide a tool to classify subjective experiment (and all its results) as
either consistent or inconsistent. Additionally, the tool identifies stimuli having
irregular score distribution.
The approach is based on treating subjective scores as a random
variable coming from the discrete Generalized Score Distribution (GSD).
The GSD, in combination with a bootstrapped G-test of goodness-of-fit,
allows to construct $p$-value P--P plot that visualizes experiment's
consistency.
The tool safeguards researchers from using inconsistent subjective data. In 
this way,
it makes sure that conclusions they draw and tools they build are more precise
and trustworthy.
The proposed approach works in line with expectations drawn solely
on experiment design descriptions of 21 real-life multimedia
quality subjective experiments.
\end{abstract}


\section{Introduction}
Any system built with an end-user in mind should provide excellent experience
for that user. In telecommunications we say that the system should provide
high Quality of Experience (QoE). Therefore, numerous works focus
on QoE optimization in various contexts. Many target quality improvement
by better network resources utilization
\cite{Bhat2018Improving, Qin2019Quality-Aware}. Some go beyond that
and use the QoE to optimize parameters like power consumption of mobile
device displays \cite{Yan2015Exploring} or to improve user-perceived quality
when watching 360$^{\circ}$ videos.
All this research is based on the concept of measuring the QoE
through subjective experiments (i.e., by asking selected end-users
about their perception of quality of a service, system or a single multimedia
material of interest). The key takeaway here is that subjective experiments
are necessary to assess the QoE.

Subjective experiments are not precise measuring systems. 
People are able to differentiate a limited number of intensities of
a given stimulus \cite{Miller1957Magical}. Additionally, they are not
perfectly consistent with their actual perception when formulating their
judgement. We thus classify collecting subjective scores as a noisy process.
There are recommended ways to analyse subjective scores and correct
for typical errors. These include subject bias removal or discarding study
participants with opinion not correlated with the general opinion of others
(cf. Rec. ITU-T P.913). However, there is no tool for classifying the whole
subjective experiment as either consistent or inconsistent. Such a tool could
protect researchers and practitioners from using erroneous data and reaching
ill-founded conclusions. Since discarding all the results at once is expensive
(because it translates to re-organizing the whole experiment again), it
would be also helpful to have a tool that can point to individual problematic
stimuli (e.g., images or videos). Hopefully, such a tool could point to
stimuli for which study participants do not agree (or express any other unwanted
behaviour). By investigating these stimuli in greater detail and, potentially,
discarding them, the general data consistency would improve (not necessitating
the whole experiment to be discarded).

The main contribution of this paper is a tool classifying the whole subjective
experiment as either consistent or inconsistent. The tool also points to specific
stimuli with irregular (also referred to as atypical) score distribution.
The overall consistency of an experiment is visualised using $p$-value P--P plot
(cf. Fig.~\ref{fig:teaser}).

There are four goals we address with this paper. First, we introduce the
new data analysis method. Second, we present which score distributions the
method classifies as typical (and thus consistent). Third, we show how the
method operates on real-life subjective data. At last, we hope to convince
the community that $p$-value P--P plots are a useful data analysis tool.

To validate the proposed method we use results from 21 real-life
subjective experiments. These test the quality of the total of 4,360 stimuli and
contain 98,282 individual scores.
Among the stimuli are: videos (with and without audio), 
images, and audio samples. 
The analysed experiments cover different experiment 
designs and have a varying number of scores per stimulus (between 9 and 33)
with most of them having the typical number of 24 scores per stimulus.

Having validated our method on the real-life data sets we believe it can be used
to assess consistency of many different subjective experiments. Furthermore,
method's ability to point to specific problematic stimuli eases data analysis
and has a potential to provide new per-stimulus insights (e.g., indicate that
a given image is especially difficult to score).

Next section highlights related work. Section~\ref{sec:model} describes
the theoretical background underlying $p$-value P--P plot and the proposed
method. In
Section~\ref{sec:gsd_as_an_experiment_consistency_check_tool} $p$-value 
P--P plot is explained. In Section~\ref{sec:analysis} the real-life data sets are
described
and analysed using our method. At last, Section~\ref{sec:conclusions} concludes the paper.


\section{Related Work}
\label{sec:relatedwork}
Subjective scores analysis is a broad topic. It is considered by numerous publications.
This includes ITU standards and, among these, the ITU-T Rec. P.1401. Although
focused on objective quality algorithms, it gives important guidelines regarding
good practices related with subjective data analysis in general.
In the similar vein, the work of Brunnström and Barkowsky~\cite{Brunnstrom2018}
looks into appropriate sample sizes for subjective experiments. In particular, they
devise a methodology to select the number of participants necessary to measure
a certain MOS score difference between stimuli.
An important observation here is that both mentioned publications rely on MOS
score analysis. They also implicitly assume correctness (also referred to as consistency)
of data stemming from subjective experiments. Our work complements the toolkit
introduced by these works allowing to check whether subjective experiment
consistency is a valid assumption. Additionally, our method gives per-stimulus information
(which is in contrast to MOS-only approach).

Search for analysis methods going beyond the MOS is present in few existing
works. One of these is \cite{Hossfeld2016}, where authors explore several
measures related to user behavior and service acceptance. Both \cite{Janowski2009}
and \cite{Seufert_Fundamental_2019} go beyond the MOS by targeting score
distributions. They propose to model the probability of each
single score, rather than focusing on just the mean. Finally, works like
\cite{Fiedler2010} operate on higher levels of abstraction and propose
a mapping between QoE and Quality of Service (QoS). With our work we also
go beyond the MOS, but do not take into account the QoE--QoS mapping.

Not many publications analyse subjective scores by focusing on the answering
process. An important exception here is \cite{Hossfeld2011}. There, the relation
between the MOS and the standard deviation of opinion scores is studied. Authors
of \cite{Hossfeld2011} introduce the HSE $\alpha$
parameter.\footnote{In the original paper
the parameter is labelled as SOS $\alpha$. Since SOS
is traditionally used to denote the standard deviation of opinion
scores, not wanting to introduce ambiguous name we refer to
the parameter using first letters of surnames of its authors.}
They argue that
it can be used to both comprehensively summarize subjective experiment results
and check their consistency.
Another interesting approach to subjective scores consistency is shown
in \cite{Hossfeld2018}. There, various confidence intervals for MOS values are
analysed.
Our approach is to analyse consistency on a per-stimulus basis and
is thus distinct both from \cite{Hossfeld2011} and \cite{Hossfeld2018}.

Our method is based on the concept of subject model. The idea was first presented
in \cite{Janowski2015} and later in \cite{Li2017}. The subject model further
extends the toolkit for subjective data analysis 
\cite{Tobias_no_silver_bullet_2017, Kumcu_2017_Performace_of_four}.
It also helps make existing analyses more precise \cite{Improve_analysis_2018, 
Freitas_2018_Performance}. The initial shape of the subject model
is modified in \cite{Pablo_AMP_2019, Tasaka_2017_Bayesian_Hierarchical}
and \cite{User_Model_for_JND_2018}. The derived models helped discover new phenomena
(e.g., observing that content has a significant influence on both
the standard deviation and mean
of opinion scores \cite{Pablo_AMP_2019}).
The analysis presented in our paper is based on a yet different
subject model that was introduced in \cite{janowski2019generalized}.

We do not compare our method to similar methods present
in other fields that also deal with quality assessment.
This is because they require different set of data
than what is usually gathered in QoE subjective experiments.
One example is a tool used by the food industry
called ``Panel Check'' \cite{Romano_Correcting_2008, 
Tomic_Visualization_2007}. Its main drawback is the
assumption of no tied answers. When using the 5-point scale
(typical for QoE experiments) this assumption is difficult to satisfy.
Another example is the signal detection
theory (SDT), extensively used in psychology \cite{Maniscalco_SDT_2012, 
Fleming_Self-evaluation_2017}. Again, it requires measurements
that are often not available in QoE experiments. Specifically, both
the recognition score and quality score are needed. In the typical
QoE experiment only the latter is provided.

P--P plots we use are a member of the family of plots referred to as
probability plots. The second member of this family are more popular
Q--Q plots. Both types of plots were introduced by Wilk
and Gnanadesikan in 1968 \cite{Wilk1968}. Their general purpose
is to compare two sets of data. Commonly, this is used to juxtapose a set
of observations with a theoretical model. The probability plots are used
throughout various disciplines, including astrophysics \cite{Lupu2012} and
landscape and urban planning \cite{DeMontis2012}.
     

\section{Subjective Score as a Random Variable}
\label{sec:model}
Scores (also referred to as answers)
collected for a single stimulus in a subjective experiment can be 
modelled by a probability distribution. Expressing an answer as a random variable has 
numerous advantages. One is the ability to formally describe typical and
atypical distributions. In this section we start from a discussion about
possible score distributions and then describe a discrete distribution 
which is used to model these.

\subsection{Typical or Atypical?}
\label{ssec:typical_or_atypical}
People do not usually use numbers to describe service quality. Instead, they
voice their judgement as single words (e.g., ``ok'') 
or more complex verbal explanations. Nevertheless, their opinion can
typically be mapped to selected categories.
Therefore, 
the most popular scale used in subjective experiments is a categorical five-point
scale (1---``Bad,'' 2---``Poor,'' 3---``Fair,'' 4---``Good'' and 5---``Excellent'').
We also use this scale, but note that the
analysis can be extended to other scales as well (see \cite{janowski2019generalized}
for the explanation).

Even in an ideal world a subject (i.e., study participant) using a discrete scale 
naturally generates scores with some degree of randomness.
For example, if stimulus quality is not satisfactory enough 
to obtain rating ``Good,'' but also not sufficiently bad to obtain rating ``Fair,'' an
ideal subject would not use the same answer each time the same stimulus is shown.
In other words, they would alternate between ratings ``Good'' and ``Fair.''
This phenomenon alone makes score distribution analysis both important and 
challenging. It also justifies why it would be valuable to be able to discern between
typical and atypical score distributions. Importantly, by typical we mean
score distributions that we would
expect to observe in a consistent subjective experiment. On the other hand,
by atypical we mean distributions appearance of which would be justifiable only by
referring to some external factors
influencing the scoring process---e.g., a bias \cite{Zielinski2008}.

We now analyse typical and atypical score distributions. 
To do so we divide the two classes (i.e., typical
and atypical) into more granular classes.
We name each class and shortly describe it. Additionally, next to the name
of each class we show an exemplary sample of scores corresponding to this class.
This sample is presented as $[k_{1}, k_{2}, k_{3}, k_{4}, k_{5}]$, where 
$k_{j}$ denotes the number of answers of category $j$. 
To simplify the discussion we assume we are always considering a single
stimulus having the true quality $\psi=2.5$.\footnote{By true quality we 
understand a non-directly-observable true quality of a stimulus. We use $\psi$ as 
defined by SAM (Statistical Analysis Methods) of VQEG (Video Quality Expert 
Group) \cite{Janowski2019NotationFS}.} Additionally, we assume
this stimulus is assigned 24 scores. In other words, $\sum_{j=1}^{5}k_{j}=24$.
At last, it is worth pointing out that the Mean Opinion Score (MOS) for
the exemplary sample should be close to 2.5 (i.e., 
$\sum_{j=1}^{5}j \cdot k_{j}/n\approx 2.5$).

These are score distributions that we treat as typical:
\begin{itemize}
	\item \textit{Perfect} $[0, 12, 12, 0, 0]$---represents the 
	most stable answers we can image. All answers are as close to $\psi$ 
	as possible. In a typical subjective experiment subjects do not agree 
	perfectly with each other so this is not the most common class.
	\item \textit{Common} $[2, 11, 9, 2, 0]$\footnote{Note that this time the 
	sample mean is not 2.5. Since $\psi$ is the true and hidden parameter, not 
	always the observed sample has the mean of 2.5.}---shows answers spread 
	around $\psi$. This class is common in real-life subjective experiments.
	\item \textit{Strongly spread} $[5, 8, 6, 4, 1]$---it is still typical but 
	starts to contain a surprisingly large spread of scores. The border line between 
	\textit{common} and \textit{strongly spread} depends on the experiment, 
	stimulus difficulty, subject pool, and possibly other factors.
\end{itemize}

To some extent it is more interesting to consider score distributions that are 
atypical. Here is a list of these that we treat as such:
\begin{itemize}
	\item \textit{Random answer(s)} $[1, 11, 11, 0, 1]$---represents single answers
	appearing away from the majority of other answers.
	\item \textit{Bimodal} $[9, 0, 10, 5, 0]$---represents a
    mixture of very different opinions. Lack of answers ``Poor'' in 
	the sample shows that we have non-uniform groups of subjects.
	Potentially, some 
	of them are experts and other are na\"ive observers. 
	In such a case we should not analyze the scores as though they were coming
	from a single group of observers.
    For example, stating that this stimulus has the true
	quality of 2.5 would be incorrect. This is because for one group the quality is 
	``Bad'' and for the other it is a little bit better than ``Fair.'' 
	\item \textit{Sudden cut-off} $[3, 7, 14, 0, 0]$---does not represent
	an obvious error. We could imagine that this exemplary sample 
	is related with a stimulus of the quality slightly below rating ``Fair.''
	Nevertheless, we should still observe more 
	``Poor'' scores and less ``Fair'' ones. The lack of ``Good'' answers
	also seems unusual, especially since there are so many ``Fair''
	ratings. Importantly, all of this can be due to chance since 24 scores
	may not be enough to accurately represent the true underlying distribution.
	Another explanation is presence of some specific disturbance in the voting 
	process.
	\item \textit{Hate or love} $[11, 3, 2, 3, 6]$---we doubt that 
	such distribution can be observed in subjective experiments.
	\item \textit{Wrong MOS}---any sample with the mean value very far from 2.5.
	 Significantly, this problem can be detected only if we know the true quality.
\end{itemize}

\subsection{Subject Model}

In order to distinguish between typical and atypical score distribution we need a subject 
model\footnote{Probably ``scoring model'' would be a better name. Sill, since it
has already been called subject model in the literature \cite{Janowski2015}
we stick to this convention.} accepting typical
samples and excluding the atypical ones. It should have as few 
parameters as possible. The smallest number of parameters is two. One is the 
true quality $\psi$, while the other describes the answers spread $\theta$.
We consider a 
model where an answer for a stimulus is a random variable $U$ drawn from a 
distribution:
\begin{equation}
U \sim F(\psi, \theta),
\end{equation}
where $F()$ is a cumulative distribution function. 

We could use two types of models: (i) continuous (proposed in
\cite{hossfeld2020qos}, \cite{Janowski2015} and
\cite{Li2017}) and discrete (proposed in 
\cite{janowski2019generalized}). Since the discrete model better fits data from
multiple subjective experiments 
\cite{janowski2019generalized} we use it as the basis of the presented method.
The continuous model could be used as well.

We refer to the discrete model as \textit{discrete distribution}. We use this
term (instead of \textit{discrete model}) purposely. We are looking for a discrete 
distribution which is a function of two parameters. A distribution
satisfying our 
constraint is the Generalized Score Distribution (GSD) described in 
\cite{janowski2019generalized}.
Since the complete description of the distribution is lengthy
we refer the reader to \cite{janowski2019generalized}. Here, we only describe 
GSD properties that are relevant for this work.

For a distribution with limited support the variance is limited as well. 
Moreover, if a distribution is discrete the lower and the upper values of the 
variance are limited and depend on the mean value.
For a five-point scale and 
the mean ($\psi$) equal to 1.5, the smallest possible variance is obtained for 
50\% answers 1 and 50\% answers 2.
In contrast, if only scores 1 and 5 are used we 
obtain the maximum variance. The obtained variance limitations (for $\psi = 
1.5$) are: $V_{\mathrm{min}}(1.5)=0.25$ and $V_{\mathrm{max}}(1.5)=1.75$. For a 
different $\psi$, for example $\psi=3$, the minimal and maximal values are: 
$V_{\mathrm{min}}(3)=0$ and $V_{\mathrm{max}}(3)=4$, respectively.
Connection 
between the variance and the mean makes the analysis difficult 
since the variance cannot be analysed without knowledge of the mean.
One way to deal with this problem is to use the HSE $\alpha$ 
parameter \cite{Hossfeld2011}.

GSD distribution parameters are the true quality $\psi$
(defining the stimulus quality) and $\rho$ (describing answer spread with 
answers closer to $\psi$ if $\rho$ is closer to 1). $\rho$ is limited to the
$(0, 1]$ interval, regardless of the $\psi$ value. This addresses the previously
mentioned problem of the variance-mean dependency.
Tab.~\ref{tab:GSDExample} presents the GSD distribution for various
$\rho$ values.

\begin{table}[!t]
    \renewcommand{\arraystretch}{1.3}
    \caption{Presentation on How the $\rho$ Parameter Influences the
    Score Distribution Modelled With GSD When $\psi=2.1$.}
    \label{tab:GSDExample}
    \centering
	\begin{tabular}{l|c|c|c|c|c|} 
		\tikz{\node[below left, inner sep=4pt] (def) {$\rho$};%
			\node[above right,inner sep=1pt] (abc) {\bfseries Score};%
			\draw (def.north west|-abc.north west) -- (def.south east-|abc.south east);} & \bfseries 1 & \bfseries 2 & \bfseries 3 & \bfseries 4 & \bfseries 5 \\ \hline
		0.95 & 	0.061 & 0.795 & 0.130 & 0.013 & 0.001 \\ \hline
		0.88 &  0.145 & 0.647 & 0.173 & 0.032 & 0.003 \\ \hline
		0.81 & 	0.230 & 0.500 & 0.215 & 0.050 & 0.005 \\ \hline
        0.72 &  0.317 & 0.370 & 0.222 & 0.078 & 0.013 \\ \hline
            0.61 &  0.394 & 0.285 & 0.184 & 0.100 & 0.037 \\ \hline
            0.38 &  0.532 & 0.153 & 0.108 & 0.096 & 0.111 \\ \hline
	\end{tabular}
\end{table}

All the typical score distribution classes described in
Section~\ref{ssec:typical_or_atypical} (i.e., \textit{perfect}, \textit{common}, and 
\textit{strongly spread}) come from the
GSD distribution. On the other hand, the atypical classes (i.e., \textit{random 
answer(s)}, \textit{bimodal}, and \textit{sudden cut-off}) are not part of GSD. 
The only atypical class that comes from the GSD distribution is \textit{hate or 
love}. This class is part of GSD since it is a generalization of the
\textit{strongly spread} class.
Therefore, we have to manually set a boundary score spread that
would still correspond to the \textit{strongly spread} class and not 
to \textit{hate or love}.
We leave for future research the task of quantifying this boundary condition.
Significantly, our data analysis have shown that \textit{hate or love} samples are rare.

Typical score distribution classes are from the GSD distribution
and atypical are not. We can thus claim that GSD reflects intuitions of the
subjective testing community developed over the years through practical
encounters with subjective data. As such, it shows which stimuli have score
distributions that would be counter-intuitive to practitioners and which follow their
intuitions and experiences. Naturally, GSD is a model that tries to simplify the
complex nature of reality. This means it should not be used as the ultimate
measure of subjective data consistency. Instead, it should be juxtaposed with
other consistency measures.

To check whether a sample of scores follows the
GSD distribution we have to first estimate the GSD parameters for the sample
and then validate whether the sample comes from GSD with
these estimated parameters. This is presented in next section.
The section also shows how to extend this reasoning to multiple samples.

\section{$p$-Value P--P Plot}
\label{sec:gsd_as_an_experiment_consistency_check_tool}

The main contribution of this paper is a new, for the QoE community, tool assessing 
subjective experiment consistency that also detects which stimuli should be
analyzed in greater detail.
The algorithm, philosophy behind it
and practical issues related with its usage are described
in this section. 

We start from the overview of the philosophy behind  the proposed methodology: 
\begin{enumerate}
	\item A consistent experiment contains stimuli with typical score distributions.
	\item Typical distributions are described by the GSD distribution.
	\item For each stimulus we can estimate the probability of whether its
	score distribution comes from the GSD.
	\item If for many stimuli this probability is low, we  should analyze the data in
	greater detail.
	\item $p$-Value P--P plot reveals when the detailed analysis is needed
	and when the experiment can be treated as consistent.
\end{enumerate}

\begin{figure}
	\centering
	\begin{tikzpicture}[->,>=stealth']
	\node (data) [state, color = blue] {\parbox{2.5cm}{\centering Subjective 
	data }};
	
	\node (dist) [state,    	
	yshift=-0.9cm, 		
	right of=data, 	
	node distance=2cm, 	
	anchor=center] {\parbox{2.5cm}{\centering GSD}};
	
	\node (par) [state,    	
	below of=dist, 	
	node distance=1cm, 	
	anchor=center] {\parbox{2.5cm}{\centering $(\psi, \rho)$}};
	
	\node (prob) [state,   
	below of=par, 	
	node distance=1.1cm, 	
	anchor=center] {\parbox{2.5cm}{\centering Expected frequencies}};
	
	\node (obs) [state,    	
	yshift=-1.9cm,
	left of=data, 	
	node distance=2cm, 	
	anchor=center] {\parbox{2.5cm}{\centering Observed frequencies}};
	
	\node (chi2) [state,    	
	yshift=-2.2cm, 		
	right of=obs, 	
	node distance=2cm, 	
	anchor=center] {\parbox{2.5cm}{\centering G-test of\\ goodness-of-fit}};
	
	\node (pval) [state,    	
	below of=chi2, 	
	node distance=1cm, 	
	anchor=center, color = red] {\parbox{2.5cm}{\centering $p$-Value}};

	\path 	(data)	edge[bend left=10]  (dist);
	\path (dist) edge (par);
	\path (par) edge (prob);
	\path (data) edge[bend right=20] (obs);
	\path 	(prob)	edge[bend left=10]  (chi2);
	\path (obs) edge[bend right=20] (chi2);
	\path (chi2) edge (pval);
	\end{tikzpicture}
	
	\caption{Algorithm to test the goodness-of-fit. The pipeline starts at 
	the blue ``Subjective data`` block. The output is a $p$-value (red box), 
	which we use to verify the null hypothesis that a sample comes from the GSD 
	distribution.}
	\label{fig:test}
\end{figure}
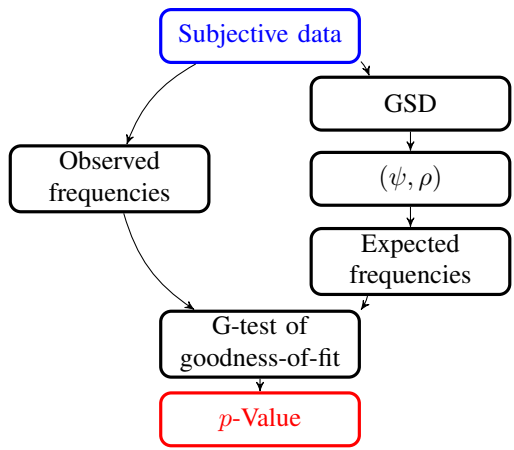

In order to check if any assumed distribution fits specific data
we have to perform a two-step procedure.
In the first step, distribution parameters are estimated for an observed 
sample. In our case, we treat scores assigned to a single stimulus as a single 
sample. The second step is to use a goodness-of-fit (GoF) test
to see how well the selected distribution (with the estimated parameter values) describes
the sample. The GoF test returns a $p$-value, which states how likely it 
is to observe the sample, assuming it comes from the considered distribution. 
Fig.~\ref{fig:test} visualizes the procedure. 
%

The GoF test we use is a standard likelihood ratio approach
called G-test (cf. Sec. 14.3.4 of \cite{Agresti}). 
Since our sample sizes are usually small we do not use the asymptotic 
distribution for calculating the $p$-value (as is the case for the popular 
$\chi^2$ test of GoF---cf. Ch. 11 of \cite{Nas2010_Basic}). 
Instead, we estimate the $p$-value 
using the bootstrapped version of G-test. The approach is described
in our GitHub repository \cite{repo}.
Broader theoretical considerations are in \cite{Efron}.

The analysis shown in Fig.~\ref{fig:test} generates as many $p$-values as 
there are stimuli in the experiment. 
With multiple $p$-values the analysis is more involved than usually
\cite{Brunnstrom2018}.
Since we 
are not interested in detecting which stimulus is not from the GSD distribution,
but rather in concluding about the experiment consistency as a whole, 
the $p$-value P--P plot is our main analytical tool
\cite{Schweder1982_Plots}.\footnote{To grasp
why $p$-value P--P plot is the better choice here we suggest to read
\cite{rosenblatt2013practitioners}.} 

$p$-Value P--P plot (see Fig.~\ref{fig:teaser} or Fig.~\ref{fig:pvalue_qqplots}) presents
a one dimensional sample of observed quantities (in our case, a vector of $p$-values)
in relation to a theoretical, expected probability distribution (in our case, the uniform
distribution spanning the range from 0 to 1). The plot is based on empirical cumulative
distribution function (ECDF) of the observed sample ($y$ axis) and
CDF of the expected probability distribution ($x$ axis). Happily, since $p$-values
are in the range from 0 to 1 and our expected uniform distribution is defined over
the same range, values on the $x$ axis not only correspond
to the expected CDF but also to the observed $p$-values.

For a Simple Hypothesis,\footnote{Any hypothesis that 
specifies population distribution completely.} such as the one stating 
that the sample is from the standard Gaussian distribution,
the expected shape of the $p$-value P--P plot is a straight line ($x 
= y$). This is because we expect $p$-values to follow the uniform 
distribution \cite{Robinson2014_How}. 
Still, deviating from the $x = y$ line
does not necessarily prove that the obtained $p$-value distribution is odd.
Slight deviations are tolerable due to the random nature of the underlying process.
Importantly, the assumption about the $x = y$ line is not valid if we consider a Composite
Hypothesis.\footnote{Any hypothesis that does not 
specify population distribution completely.} One such example is a
hypothesis stating that a sample is from the GSD distribution with unknown
parameters. In this case, the expected shape of the $p$-value P--P plot is
best described by a set of points falling below the $x=y$ line
(see Fig.~\ref{fig:ex_consistent_exp}). Only if the points significantly exceed
the line can we reject our hypothesis.


To quantify what it means to significantly exceed the $x=y$ line we use
the following procedure. 
For each considered $\alpha$ (i.e., a specific value on the $x$ axis of 
$p$-value P--P plot) we change each $p$-value to 0 or 1. 
We do it by checking whether it is larger (0) or smaller (1) than $\alpha$. 
Now, the whole experiment is changed to a vector of zeros and ones with its 
length equal to the number of 
stimuli. The question under investigation is: ``is the number of stimuli with
small $p$-value significant?'' (Note that ``small'' is defined by $\alpha$.)
This can be understood as hypothesis testing verifying the null hypothesis 
stating that the probability of finding a stimulus with $p$-value
below $\alpha$
is smaller than $\alpha$. This is a classical problem with 
the solution given by: 
\begin{equation} \label{eq:p}
\hat{\alpha} > \alpha + z_{1-\beta}\sqrt{\frac{\alpha(1 - \alpha)}{n}},
\end{equation}
where $\hat{\alpha}$ is the observed proportion of $p$-values smaller than 
$\alpha$ (i.e., the $y$ axis value on $p$-value P--P plot), $\alpha$ is the conjectured proportion of such stimuli, $z_{\gamma}$ is the quantile of order $\gamma$ for
the standard normal distribution and $\beta$ is 
the significance level of this test. (We use 5\% and hence
$z_{1-\beta}=1.64$.) Importantly, if Eq.~(\ref{eq:p}) is
satisfied we reject the null hypothesis and say that around $\hat{\alpha} - 
\alpha$ stimuli truly have their $p$-value below $\alpha$. 
Substituting the LHS of Eq.~(\ref{eq:p}) by $f(\alpha)$ and equating both sides
we get a formula for drawing a line spanning the whole range of $\alpha$ values
(from 0 to 1). This is the black line visible in Fig.~\ref{fig:teaser} and
Fig.~\ref{fig:pvalue_qqplots}. When points on $p$-value P--P plot significantly
exceed this threshold line we say that an experiment is inconsistent.
In other words, we reject our composite hypothesis stating that
score distributions of stimuli in an experiment come from the GSD distribution.

As stated in the previous paragraph Eq.~\eqref{eq:p} is used to draw the 
theoretical threshold line in $p$-value P--P plot. Now, if data
points are crossing this line (i.e., falling above it) or are close to it we 
should analyse all stimuli with $p$-values smaller than $\alpha$ of the 
crossing point. Note that $\alpha$ of the crossing point is found by looking
at the $x$ axis value on the P--P plot for which the data crosses the theoretical
threshold. 
Analysing the stimuli with $p$-values smaller than the crossing point it is 
important to remember that it is natural to
observe a fraction of stimuli with $p$-value below $\alpha$ as high as
$\alpha$. In other words, even if we draw the scores
from the GSD itself, still the fraction of stimuli with $p$-value below $\alpha$
could be close to $\alpha$. In any case we are not allowed to simply remove all 
stimuli with $p$-value smaller than $\alpha$. Instead, we should analyse their 
score distribution one-by-one and look for specific problems.
Section~\ref{ssec:case_study_on_real_data} demonstrates
such an analysis.

Describing the above algorithm from the practical perspective let us consider 
generating 160 samples, each time drawing 24 values from the GSD distribution. 
This way we simulate a subjective 
experiment with 160 stimuli, each assigned 24 scores.
Since scores come from the GSD distribution we know that each stimulus
has a typical score distribution.
We now apply the procedure from Fig.~\ref{fig:test} to get $p$-value
for each stimulus.
This generates a range of $p$-values, some of which are small
(e.g., smaller than 0.05). These small $p$-values appear even though we
know the input data is generated from typical score distributions only.
Hence, the decision of whether to classify some real 
experiment, as a whole, as inconsistent cannot be based on observing few small 
$p$-values. 
Also, just a single stimulus with low $p$-value cannot be labelled as
odd \cite{Robinson2014_How}.
Instead, the proportion of $p$-values (relative to the total number of stimuli) 
smaller than certain $\alpha$ should be analysed---cf. Eq.~\eqref{eq:p}.

Our simulation study proved that if the data comes from the GSD distribution
points on $p$-value P--P plot fall below or spontaneously coincide with the
threshold line.
Since this holds for the region of $p$-values 
that is critical for drawing conclusions (i.e., for $\alpha<0.2$), we can follow the
standard recommended analysis from \cite{Schweder1982_Plots}. 

\paragraph{GSD Parameters Estimation}
A practical issue related with
$p$-value P--P plot is GSD parameters estimation.
GSD has a complicated form of its likelihood function. Thus, it is not
possible to provide analytical solution for parameters estimation.
To simplify the estimation we pre-calculate probabilities of each
score for a grid of $\psi$ and $\rho$ values. Specifically, we use
399 values of $\psi$, spanning the interval $[1.01, 4.99]$, with the step
of 0.01; and 400 values of $\rho$, spanning the interval $[0.0025, 1]$, with
the step of 0.0025.
The pre-calculated grid accelerates the estimation
process and protects us from any potential difficulties that arise
when using dynamic parameter optimization.
Using simulation studies we have validated that the proposed
estimation method works as expected. In other words, we are able to
recover correct GSD parameters from synthetic data that is itself generated
from the GSD.

\section{Subjective Data Analysis}
\label{sec:analysis}
This section puts our idea into practice. We first describe
six real-life subjective studies we use. The descriptions highlight
reasonable expectations about studies consistency. We then
verify these expectations by applying our method.

\subsection{Data Sets}
\label{ssec:data_sets}

To check the practical distribution of subjective scores we use data from 21 subjective experiments.
The data comes from six studies representing various stimulus types: (i) VQEG HDTV Phase I \cite{HDTV_Phase_I_test} (six experiments; video-only), (ii) ITS4S \cite{ITS4S} (two experiments; video-only), (iii) AGH/NTIA \cite{Janowski2014, AGH_NTIA_14-505} (one experiment; video-only), (iv) MM2 \cite{Pinson2012} (ten experiments; audiovisual),
(v) ITS4S2 \cite{ITS4S2} (one experiment; image), and (vi) ITU-T Supp. 23 \cite{itutsupp23} (one experiment; speech). 
This gives a total of 98,282 subjective scores. 

All the experiments use the 5-level Absolute Category Rating (ACR) experiment design. 
Importantly, all the data we use is publicly accessible. Please refer
to the references provided for each study or go to our
repository \cite{repo}
to download a single CSV file with all the results combined (and put into the 
tidy data format~\cite{Wickham2014}).
 
\paragraph{VQEG HDTV Phase I} All the six experiments from this study took 
place in a controlled environment. Test participants were screened for normal
vision acuity and normal colour vision. Only scores of those testers who 
correlated well with the average opinion were kept. Impairments included 
compression and lossy transmission. All these suggest that it is fair to
expect consistent results.

\paragraph{ITS4S}
Utilizing the unrepeated scene design this study
has a potential to strongly emphasize personal preference differences between 
participants. With no strict vision acuity and colour blindness testing the 
study has a higher chance to include unreliable subjects. 
Furthermore, it uses content that can trigger strong 
emotional response. 
Topping this with the scores originating from a mixture of experts
and na\"ive subjects, we expect this data
to be less consistent than that of HDTV Phase I. Still,
the study setting was not completely uncontrolled as participants were
seated in a lab-like environment. They were briefed and
trained as well. This suggests better scores consistency than
that of crowdsourced studies. Due to experiment design choices (other than those described above),
we expect the first experiment to be more consistent than the 
second one. For one reason, the second experiment allowed
participants to take the test simultaneously in groups of ten.

\paragraph{AGH/NTIA}
Non-standard procedures in this test are: lack of 
screening for normal vision acuity and colour vision; distance to the screen was not 
strictly controlled; two testers received intentionally erroneous instructions 
and one tester was a video quality expert. However, the study generally
complied with Rec. ITU-T P.910, recruited testers through a temporary job 
recruitment agency and investigated compression as the only distortion. We 
expect this study to be less consistent than HDTV Phase I, but more consistent 
than other more loose experiment designs.

\paragraph{MM2}
Five out of ten experiments took place in a laboratory-like
environment. The other half took place in a less controlled 
setting.\footnote{For a detailed specification please refer to Table IV in
\cite{Pinson2012}.} All the experiments used the same set of 
audio-visual sequences. Significantly, the presence of audio might 
have increased inter-tester difference of opinion. The study investigated
compression as the only distortion source. Furthermore, all 
participants went through training and briefing. In general, we
expect MM2's laboratory experiments to be more stable than those with
the loose setting.

\paragraph{ITS4S2}
Although compliant with Rec. ITU-T P.913 this study has the
greatest potential for inconsistent results. It investigates non-standard distortions
related to consumer-grade cameras. This makes the study novel but
also means that different (than those recommended) best practices may apply.
At last, it includes content that may trigger strong emotional response.

\paragraph{ITU-T Supp. 23}
This study investigates the subjective performance
of a speech codec. Since it follows a strict and well-controlled experiment design
we expect its scores to be very consistent.
Importantly, we only utilize data from the experiment
conducted by Nortel (although the study contains three experiments in total).

\subsection{Case Study on Real Data}
\label{ssec:case_study_on_real_data}
We start from classifying each of the 21 experiments as either consistent or
inconsistent. For this purpose we use the hypothesis testing approach described 
in Section~\ref{sec:gsd_as_an_experiment_consistency_check_tool}. Selecting
the $\alpha = 0.2$ we identify four experiments as inconsistent.
Table~\ref{tab:hyp_testing} lists those, starting from the one with the smallest
$p$-value.\footnote{Please note that this $p$-value describes the
hypothesis testing outcome and not the goodness-of-fit test for a particular
stimulus.}
\begin{table}
    \renewcommand{\arraystretch}{1.3}
    \caption{The Four Lowest $p$-Values of the Test Validating the Null Hypothesis That an
	Experiment Contains Mostly Typical Score Distributions.}
	\label{tab:hyp_testing}
	\centering
	\begin{tabular}{ll}
	\toprule
	\bfseries Experiment                               & \bfseries $p$-Value \\ \midrule
	ITS4S2                                   & 0.00263   \\
	ITS4S---2\textsuperscript{nd} experiment & 0.02320   \\
	ITS4S---1\textsuperscript{st} experiment & 0.02476   \\
	MM2---IRCCyN (lab env.)                  & 0.02634   \\ \bottomrule
	\end{tabular}
\end{table}
Recalling study descriptions from Section~\ref{ssec:data_sets} only
the inconsistency of the one laboratory experiment from MM2 is a surprise.
It would be logical
to expect inconsistency in the experiments with loose settings rather than
in the one done in a lab. One potential explanation of this result is that
the MM2 study is generally consistent. However, when repeating the same
experiment ten times, the chances of randomly observing at least one
experiment being inconsistent is significant.
Furthermore, we use the MM2 data as is. This means we do not
perform any post-experimental screening of subjects before
we do our analyses. The experiment we label as inconsistent
includes data from three subjects poorly correlated with the
general opinion (with correlations of 0.49, 0.59 and 0.64).
In fact, this experiment includes two of the least correlated
subjects among all of the ten MM2 experiments (compare Fig.~2 in
\cite{Pinson2012}).
We know from experience that $p$-value P--P plot is sensitive
to poorly correlated subjects.
Thus, we hypothesize that the three outlying subjects
are the main reason our method marks the one MM2 experiment
as inconsistent.

\begin{figure}[h]
	\centering
	\subfloat[AGH/NTIA]{
	    \includegraphics[height=2.6cm]{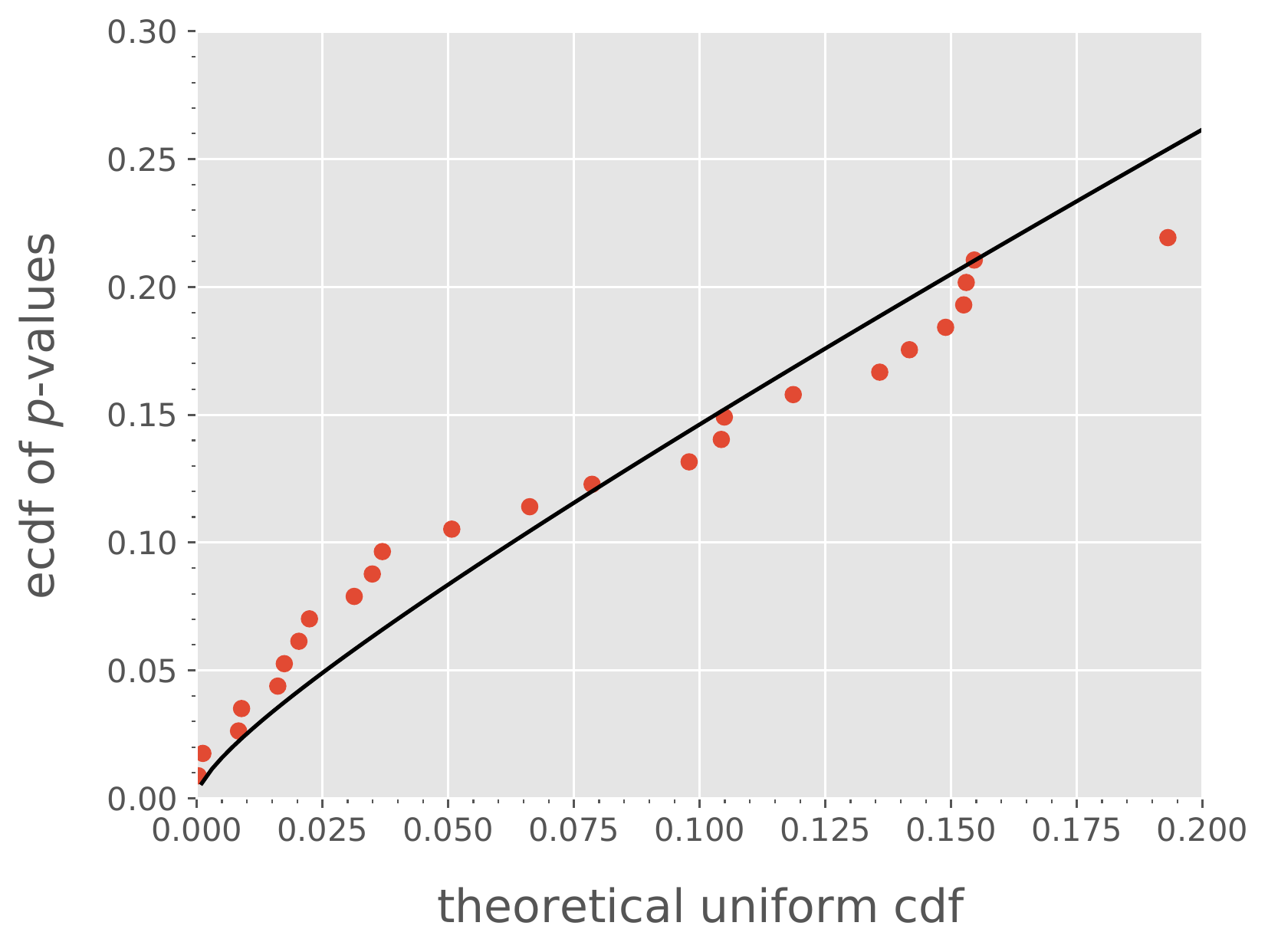}
	    \label{fig:pvalue_qqplot_agh_ntia}
	    \hfill
	}
	\subfloat[VQEG HDTV Phase I]{
	    \includegraphics[height=2.6cm]{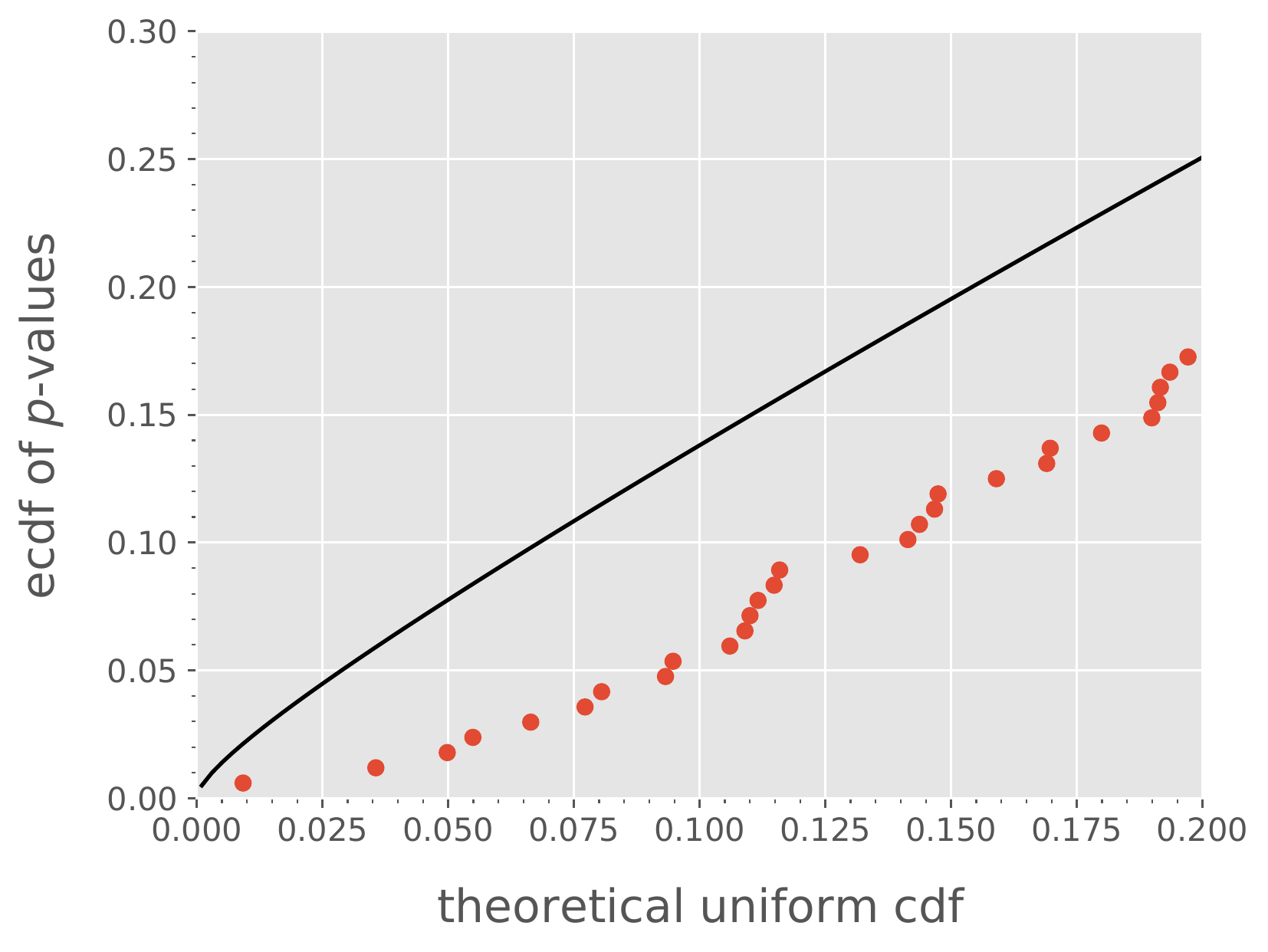}
	    \label{fig:pvalue_qqplot_hdtv}
	    \hfill
	}\\
	\subfloat[ITS4S2]{
        \includegraphics[height=2.6cm]{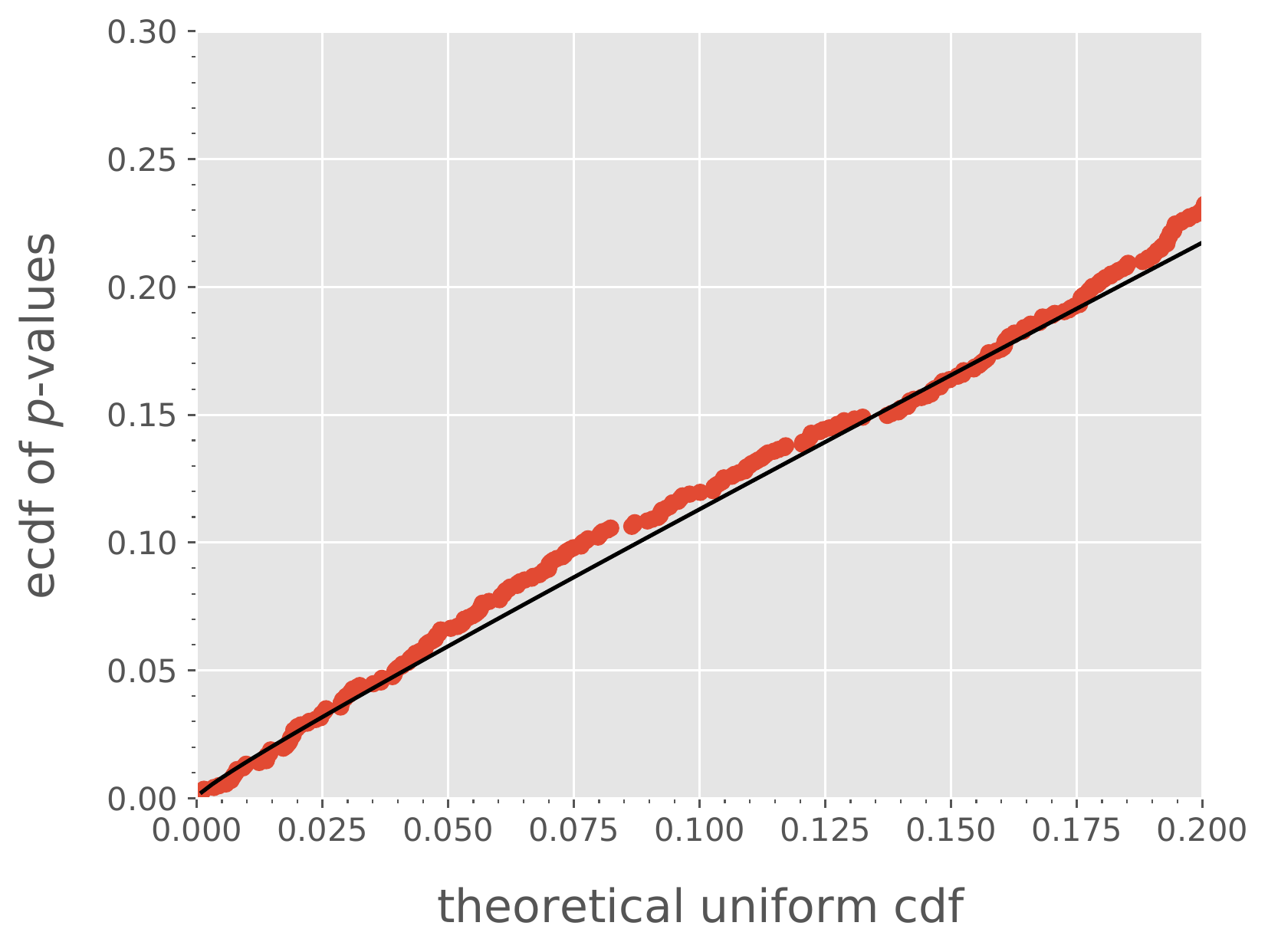}
        \label{fig:pvalue_qqplot_its4s2}
        \hfill
	}
	\subfloat[ITS4S]{
	    \includegraphics[height=2.6cm]{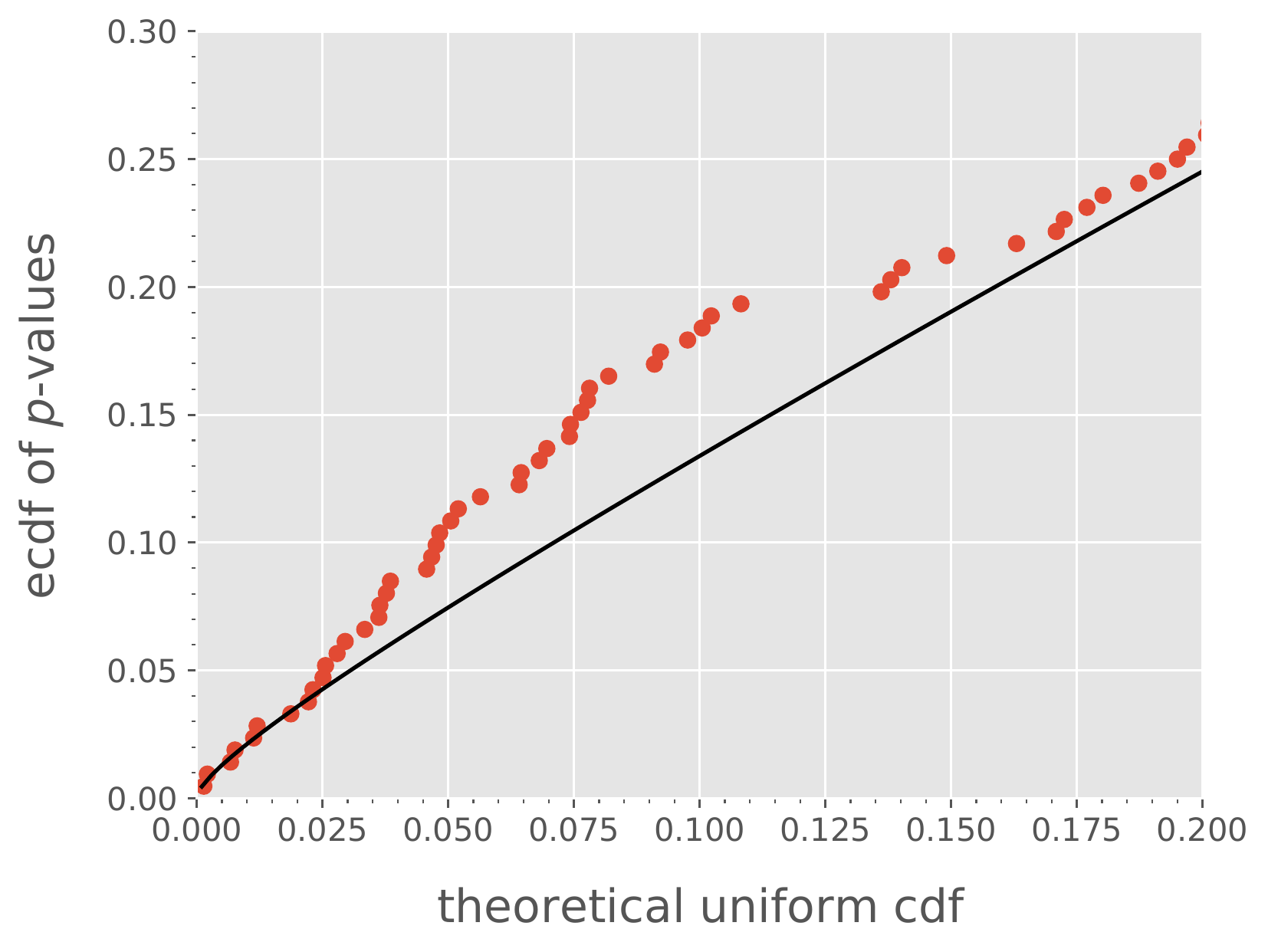}
	    \label{fig:pvalue_qqplot_its4s}
	}
	\caption{$p$-Value P--P plots for (a) the one (and only)
	experiment from the AGH/NTIA study, (b) the first experiment
	from the VQEG HDTV Phase I study, (c) the one (and only)
	experiment from the ITS4S2 study and (d) the second
	experiment from the ITS4S study.}
	\label{fig:pvalue_qqplots}
\end{figure}

Though the above analysis is based on hypothesis testing alone,
we recommend to take a look at $p$-value P--P plot of each
experiment. 
One argument for using the P--P plot is that it applies
the same reasoning as above, but for a range of $\alpha$ values.
In fact, by looking at the $p$-value P--P plot of the
AGH/NTIA experiment (see Fig.~\ref{fig:pvalue_qqplot_agh_ntia})
we see that it consistently fails the
hypothesis testing for $\alpha$ values below 0.08. This points
to at least partial inconsistency of the experiment, which is
in line with the expectations given in Section~\ref{ssec:data_sets}.

We decide not to show $p$-value P--P plot for each experiment. However,
we do provide a complete, detailed analysis of three selected
experiments. We use data from:
(i) the first 
experiment from the VQEG HDTV Phase I study (later referred to as HDTV1), (ii) 
one (and only) experiment 
from the ITS4S2 study (later referred to as ITS4S2) and (iii) the second experiment 
from the ITS4S study (later referred to as ITS4S).
Figures \ref{fig:pvalue_qqplot_hdtv}, \ref{fig:pvalue_qqplot_its4s2} and
\ref{fig:pvalue_qqplot_its4s} present $p$-value P--P plots for each experiment,
respectively. 	

By looking at the figures we can say that the experiments are sorted in 
accordance
with their consistency. With all points in Fig.~\ref{fig:pvalue_qqplot_hdtv} 
falling bellow the black line,
the HDTV1 experiment is certainly consistent. This means that we do not have to 
investigate it further.
In Fig.~\ref{fig:pvalue_qqplot_its4s2} we see points oscillating around
the black line. This suggests that the ITS4S2 experiment is neither totally
consistent nor inconsistent. In order to decide about its consistency we have to
take a close look at stimuli with low $p$-values.
At last, from Fig.~\ref{fig:pvalue_qqplot_its4s} we can quickly conclude that
the ITS4S experiment is inconsistent. 
This is because all the points fall on or above the black line. Furthermore,
the points falling above the line deviate from it significantly.
Here also we have to take a close look
at stimuli with low $p$-values. However, this time we expect to find
more such stimuli (relative to the total number of stimuli in this experiment).

Proceeding to the detailed analysis of low $p$-value stimuli we suggest
to consider all stimuli with $p$-value lower than that of the right-most point
exceeding the theoretical threshold. Please note
that we only take into account the $p$-value range from 0 to 0.2
(cf. Section~\ref{sec:gsd_as_an_experiment_consistency_check_tool} for the
explanation). This means that even if there are points exceeding the black
line with corresponding $p$-value above 0.2, we still take 0.2 as the upper
limit defining which stimuli should be analysed in detail.
By applying this rule we observe that we need to analyse all stimuli with
$p$-value below 0.2 for both ITS4S and ITS4S2.
This corresponds to 328 stimuli from ITS4S2 and
54 stimuli from ITS4S. Although the number for ITS4S2 is greater, when
analysing the figures in relative terms (i.e., comparing them to the total 
number of stimuli in each experiment), we notice that there are actually
more potentially problematic stimuli in ITS4S (25.47\%) rather than in
ITS4S2 (22.95\%).
To keep this case study concise we only take a look at five stimuli with lowest
$p$-values from both experiments. Tables~\ref{tab:lowest_pvalue_its4s}
and \ref{tab:lowest_pvalue_its4s2} present score counts for these stimuli
for ITS4S and ITS4S2, respectively. 

Studying Table~\ref{tab:lowest_pvalue_its4s} we can classify
score distribution of problematic stimuli
into one of three classes (representing a subset of the classes
described in Section~\ref{ssec:typical_or_atypical}): (i) \textit{bimodal}, 
(ii) \textit{random answer(s)} and (iii) \textit{sudden cut-off}.
\begin{table}
    \renewcommand{\arraystretch}{1.3}
    \caption{Score Count for Five Stimuli With the Lowest $p$-Value From
    the ITS4S Experiment. ID Stands for Identifier.}
    \label{tab:lowest_pvalue_its4s}
    \centering
	\begin{tabular}{@{}lllllll@{}}
	\toprule
	\multirow{2}{*}{\bfseries ID} & \multicolumn{5}{l}{\bfseries Score Count} & 
	\multirow{2}{*}{\bfseries $p$-Value} \\
	                    & \bfseries 1    & \bfseries 2    & \bfseries 3    & \bfseries 4   & \bfseries 5    &                          \\ \midrule
	a                   & 0    & 0    & 13   & 5   & 6    & 0.0014                   \\
	b                   & 2    & 0    & 0    & 9   & 13   & 0.0021                   \\
	c                   & 0    & 0    & 14   & 6   & 4    & 0.0067                   \\
	d                   & 1    & 13   & 4    & 6   & 0    & 0.0076                   \\
	e                   & 0    & 9    & 3    & 8   & 3    & 0.0113                   \\ \bottomrule
	\end{tabular}
\end{table}
\begin{table}
    \renewcommand{\arraystretch}{1.3}
    \caption{Score Count for Five Stimuli With the Lowest $p$-Value From
    the ITS4S2 Experiment. ID Stands for Identifier.}
    \label{tab:lowest_pvalue_its4s2}
    \centering
	\begin{tabular}{lllllll}
	\toprule
	\multirow{2}{*}{\bfseries ID} & \multicolumn{5}{l}{\bfseries Score Count} & 
	\multirow{2}{*}{\bfseries $p$-Value} \\
	                    & \bfseries 1    & \bfseries 2    & \bfseries 3    & \bfseries 4   & \bfseries 5    &                          \\ \midrule
	f                   & 0    & 11    & 3   & 0    & 2   & 0.0002                   \\
	g                   & 2    & 0     & 3   & 11   & 0   & 0.0004                   \\
	h                   & 1    & 2     & 1   & 12   & 0   & 0.0008                   \\
	i                   & 4    & 6     & 0   & 6    & 0   & 0.0012                   \\
	j                   & 1    & 0     & 0   & 9    & 6   & 0.0014                   \\ \bottomrule
	\end{tabular}
\end{table}
Stimuli a, d and e fall into the \textit{bimodal} class. Their score count 
suggests that there are two modes of the score distribution. 
One potential explanation for this is that there are two groups of 
participants, both expressing a significant bias. If individual biases are 
similar within the group and significantly disjoint between the groups, we 
observe the score distribution similar to the one of stimuli a, d, and e. 
It is also possible that the \textit{bimodal} distribution does not come solely 
from
characteristics of the participants as a whole. It may be that these stimuli
present content that highlights individual preferences
of the participants. In other words, participants who disagree when scoring 
these particular stimuli agree when scoring other stimuli.
It would be audacious to
suggest these stimuli should be discarded. Instead, we advise to treat this case
as a valuable insight about study participants and stimulus itself.
Our framework classifies these stimuli as a source of inconsistency, because
they do not fit the general assumptions about the score distribution. For 
example, it is difficult to believe that the MOS provides meaningful information
for such cases. The mean falls between the two modes and expresses
the general opinion of neither of the two hypothetical participant groups.

Stimulus b can be assigned to the \textit{random answer(s)} class. Here, it 
seems fairly safe to
claim that the two 1s are a result of some error. Discarding these two answers
we are still at risk of removing a genuine opinion. However, we claim that
many practitioners would agree to remove these.

Stimulus c fits into the \textit{sudden cut-off} class. We observe 
many 3s, but no 2s and no 1s. Though this is not outright wrong, it is hard 
to believe that there would be no one assigning this stimulus the score of 2.
This type of stimulus is difficult to handle. 
Even if
we discard selected scores thanks to, for example, removing poorly correlated
study participants, the sudden cut-off in the middle of the scale is likely to
remain. We advise to take a close look at stimulus' content. One hypothesis
is that this stimulus is difficult to score and people choose the middle of the
scale as the safest option conveying the ``I do not know'' message. If true,
this suggests that debriefing study participants may provide crucial insights
about how to analyse such stimuli. Another option (although usually not 
feasible) is to try to gather more scores on the stimulus. It is worth 
remembering that 24 observations may be too few to rightfully represent
the shape of the underlying score distribution.
Finally, we point out that stimulus a also seems to be of the \textit{sudden 
cut-off} class. 

Applying a similar analysis to the data in
Table~\ref{tab:lowest_pvalue_its4s2} we
note that: stimuli h and i belong to the \textit{bimodal} class, and stimuli f, 
g and j to the \textit{random answer(s)} class.
However, upon closer inspection more peculiarities become
visible. Please note that all stimuli represent data following the description 
of the \textit{sudden cut-off} class. For example, even if we discard two outlying 1s 
from the scores of stimulus g, the sudden cut-off at score 4 remains.

A careful reader will notice that the HDTV1 experiment contains
stimuli with $p$-value below the 0.2 threshold (although we stated the
experiment is consistent). Those stimuli could be analysed but we are
advising not to do so. Statistically speaking it is not
unusual to observe few low $p$-values. According to our framework
the low $p$-values are rare enough to classify the experiment 
as consistent.

This case study shows that our framework detects stimuli with atypical score
distribution. Additional information (apart from the scores) and a closer
inspection may still be necessary to decide what to do with the identified 
stimuli. Still, the method provides a handy tool reliably detecting
potential defects present in the data.


\section{Conclusions}
\label{sec:conclusions}
We introduce a new tool classifying results of a subjective
experiment as either consistent or inconsistent. The tool also highlights
stimuli with irregular score distribution. 
We show that the method works by using data from
21 subjective experiments.
Apart from the theoretical description we provide a software implementation.
To download it please go to:
\url{https://github.com/Qub3k/subjective-exp-consistency-check}.
There, we share the data set and Python
scripts used for the data analysis and a cookbook-style tutorial on
how to apply our method on arbitrary subjective data. The data set and
scripts are also provided in the auxiliary material.\footnote{This is
not valid for the arXiv version of the paper. To download the auxiliary
material please take a look at
\href{https://doi.org/10.1145/3394171.3413749}{the ACM Digital Library
entry for this paper}.}

The procedure behind our method can be summarized in two steps. First,
generate a $p$-value P--P plot for a subjective experiment and check whether
data points fall above the line defining the theoretical threshold. Second, if they
do---the experiment is potentially inconsistent and score distributions
of low $p$-value stimuli should be analysed; if they do not---the experiment is
consistent.

Though practical our method has its limitations. It cannot be directly used
to show which study participants are potential outliers. Likewise, it does not
show which observed inconsistencies are obtained by chance and which are
the result of some true underlying phenomenon (e.g., a bias influencing the scoring
process). Finally, if GSD does not cover all real score distributions (i.e., the ones
being the result of a valid subjective evaluation) we will observe false negatives
(negative meaning classifying an experiment as inconsistent). For the explanation
why this last risk seems to be small we refer the reader to
\cite{janowski2019generalized}.

We do not directly compare out method with other screening techniques since
none of them targets consistency of a whole subjective experiment. They rather focus
on discarding individual study participants or quantifying mean-variance relationship.
(Significantly, the latter is also addressed when using GSD to model score distribution.)

There are two topics that we would like to explore in our further work. First,
we would like to test if a different subject model could be used instead of the GSD model. 
In particular, we would like to test the quantized normal model~\cite{Janowski2015}.
Second, we would like to run thorough
simulation studies. This would help explore limitations of our method and show
how typical problems with subjective data influence $p$-value P--P plots.

We hope our work will serve as a useful tool for the research community and
invite everyone to test it on their subjective data.

\section*{Acknowledgment}
The authors would like to thank
Netflix, Inc.
for sponsoring this research
and especially
Zhi Li
for his support and challenging questions.
This work was supported by
the Polish Ministry of Science and Higher Education
with the subvention funds of the Faculty of Computer Science,
Electronics and Telecommunications of AGH University
and by
the PL-Grid Infrastructure.

\bibliographystyle{ACM-Reference-Format}
\bibliography{bibliography}

\end{document}